\documentclass[useAMS,usenatbib]{mn2e}

\usepackage{graphicx}
\usepackage{epstopdf}

\newcommand{\lx}{erg~s$^{-1}$}
\newcommand{\nh}{cm$^{-2}$}
\newcommand{\chandra}{{\it Chandra}}
\newcommand{\xmm}{{\it XMM-Newton}}
\newcommand{\rsun}{$R_{\odot}$}
\newcommand{\msun}{$M_{\odot}$}

\newcommand{\apj}{{\it ApJ}}
\newcommand{\apjl}{ApJL}
\newcommand{\aap}{A \& A}
\newcommand{\mnras}{{\it MNRAS}}
\newcommand{\araa}{{\it ARRA}}
\newcommand{\ssr}{{\it SSR}}

\newcommand{\porb}{{$P_{orbit}$}}
\newcommand{\pspin}{{$P_{spin}$}}

\title[Revisiting the Dynamical Case for a Massive Black Hole in IC10 X-1] {Revisiting the Dynamical Case for a Massive Black Hole in IC10 X-1 }
\author[Silas G. T. Laycock, Thomas J. Maccarone, and Dimitris M. Christodoulou]{Silas G. T. Laycock$^{1}$\thanks{E-mail:
silas\_laycock@uml.edu}, 
Thomas J. Maccarone$^{2}$ and Dimitris M. Christodoulou$^{3}$ \\
$^{1}$Department of Physics, University of Massachusetts Lowell, MA, 01854, USA\\
$^{2}$Department of Physics, Texas Tech University, Lubbock TX, 79409, USA\\
$^{3}$Department of Mathematical Sciences, University of Massachusetts Lowell, MA, 01854, USA}
\begin{document}


\pagerange{\pageref{firstpage}--\pageref{lastpage}} \pubyear{2015}

\maketitle

\label{firstpage}

\begin{abstract}
The relative phasing of the X-ray eclipse ephemeris and optical radial velocity (RV) curve for the X-ray binary IC10 X-1 suggests the He[$\lambda$4686] emission-line originates in a shadowed sector of the stellar wind that avoids ionization by X-rays from the compact object. The line attains maximum blueshift when the wind is directly toward us at mid X-ray eclipse, as is also seen in Cygnus X-3. If the RV curve is unrelated to stellar motion, evidence for a massive black hole evaporates because the mass function of the binary is unknown. The reported X-ray luminosity, spectrum, slow QPO, and broad eclipses caused by absorption/scattering in the WR wind are all consistent with either a low-stellar-mass BH or a NS. For a NS, the centre of mass lies inside the WR envelope whose motion is then far below the observed 370 km/s RV amplitude, while the velocity of the compact object is as high as 600 km/s. The resulting 0.4\% doppler variation of X-ray spectral lines could be confirmed by missions in development. These arguments also apply to other putative BH binaries whose RV and eclipse curves are not yet phase-connected. Theories of BH formation and predicted rates of gravitational wave sources may need revision.

\end{abstract}


\section{Introduction}
IC10 X-1 is a high-mass X-ray binary (HMXB) containing a compact object that accretes at $L_X >10^{38}$ \lx~from the wind of its Wolf-Rayet (WR) companion \citep{Prestwich2007}. The binary period is 35 hrs and deep eclipses are seen in the X-ray light curve. This Letter addresses the strong doubts cast by recent observations on the massive black hole (BH) interpretation for this system. A substantial downward revision of the mass of the compact object will impact research on the evolution of massive stars and BHs of all sizes (\citealt{Barnard2008, Belc2010}, \citealt{Carpano2008}, \citealt{PR2011}), long gamma-ray bursts and BH-BH sources of gravitational waves (\citealt{Bogomazov2014, Bulik2011}),  as well as the so-called Ultra-Luminous X-ray (ULX) sources (\citealt{Soria2007}, \citealt{Gladstone2009}, \citealt{Feng2011}, \citealt{Medvedev2013}). In particular, the noted preference for massive BH candidates to occur in low metallicity environments would lose its central exemplar. 

\subsection{High-Mass X-Ray Binaries}
Binary systems containing X-ray emitting compact objects accreting matter from their companion stars are thought to host both neutron stars (NSs) and black holes. The compact objects themselves are invisible due to their very small sizes and correspondingly low intrinsic luminosities. Their existence is inferred from the properties of the X-ray emission and from the doppler-derived motions of the companions as the stars orbit around their centers of mass. Spectral lines originating at the surface of the primary are observed repeatedly over the course of one or more orbital cycles to establish the amplitude of the radial velocity (RV) curve. Tidal distortion of the primary (in close systems) also leads to a periodic optical brightness modulation, known as ellipsoidal variation, which can be seen in the photometric light curve. Modeling of these variations can yield an independent measurement of the orbital period and inclination of the system. Since it is only the primary that is directly observed, the mass function is combined with independent estimates of its mass to constrain the mass of the unseen companion. These techniques have yielded a growing list (e.g., \citealt{CJ2014}, \citealt{Remillard2006}) of binary systems where the compact object mass is significantly above the theoretical limit of 2.9 \msun~ for neutron stars. Such systems are generally (although not universally) accepted to contain BHs. 

Another class of objects thought to contain stellar-mass BHs are the ULX sources (\citealt{Liu2013}, \citealt{Motch2014}). Their X-ray luminosities ($10^{39}-10^{41}$ \lx) exceed the Eddington limit for a NS by $1-3$ orders of magnitude. Most ULXs are too distant for RV or photometric studies by existing instruments to reveal their dynamical parameters, so the argument for stellar-mass BHs in these systems derives mainly from their X-ray properties. 

The X-ray binary IC10 X-1 occupies an important niche as a link between BH-HMXBs and ULXs due to its intermediate luminosity ($\sim$10$^{38}$\lx) and the fact that, at a distance of 660 kpc, it is one of the few extragalactic systems close enough to be readily observable by current generation telescopes.  The landmark work of \cite{SF} revealed that IC10 X-1 appears to have an RV-derived mass function that requires a BH of 23-32 \msun. SF noted that their RV curve had not been 
phase-aligned with the X-ray eclipses since the X-ray period was not then known to sufficient precision. Others (e.g. \citealt{Tutukov2013}) have pointed out that this caveat also applies to NGC300 X-1 which is in many regards a twin to IC10 X-1 in terms of its WR companion, similar orbital period, and X-ray luminosity. 

\subsection{Wind Ionization and Shadowing}
In the most massive and hottest stars, the stellar photospheres are highly ionized so that no suitable absorption lines can be found and emission lines are used instead. The most suitable ion species is He II because it can exist at the temperatures found in the photospheres of O stars ($>3\times 10^4$K), where it gives rise to the Pickering series of spectral lines.  Emission lines arise in low-density gas, so the region probed by the He II lines can in principle extend to many stellar radii above the surface. The density of stellar winds falls with distance $r$ approximately as $1/r^2$, so most of the equivalent width of the line should originate at lower altitudes. Provided that the wind is isotropic (or at least azimuthally symmetric), the He II line should still trace the orbital motion of the star. 


At the very high X-ray fluxes seen in high luminosity XRBs, the wind will be completely ionized out to a distance far exceeding the orbital separation of the binary. This fact has not been generally recognized as affecting the mass determinations obtained via the RV-curve method. \cite{Maccarone2014} have recently pointed out that a model introduced by \cite{vanKerkwijk1996} for Cyg X-3 might apply more widely. In this model, X-ray photoionization of the wind suppresses line emission throughout the binary orbit. Only in a small sector of the wind shielded by the primary will the gas be able to maintain a ``normal" ionization state. In IC10 X-1, the ionization parameter has a large value ($L_X/nd^2 \sim 10^3$ ) at the location of the WR star \citep{Barnard2014}. The wavelength and shape of the observed emission lines will be a convolution of the wind velocity $v_{wind}$, the orbital motion $v_{orbit}$, and the stellar rotation $v_{rot}$.  

In the limit that $v_{wind} >> v_{orbit}, v_{rot}$, the observed RV curve will have an amplitude of $v_{wind} \sin{i}$, where $i$ is the inclination of the orbital plane to the line of sight.  The phase of the resulting RV curve will be as follows: at mid X-ray eclipse, when the WR star lies between us and the X-ray source, the shielded wind sector is pointing toward the observer, as illustrated in Figure~\ref{fig:map}. The wind velocity vector then lies along our line of sight, producing the maximum blueshift of the emission line. At quadrature, when the WR star's projected radial velocity is greatest, the wind in the shielded sector is orthogonal to the line of sight, producing zero doppler shift. Finally at superior conjunction, the emitting portion of the wind is expanding away from us, leading to the maximum redshift. \cite{Hanson2000} were able to map this cool structure in Cyg X-3 from IR spectra using doppler tomography. 

The actual phase relationship between X-ray eclipses in IC10 X-1 and the RV curve of the optical counterpart has been recently explored by \cite{Laycock2015}. After constructing a new ephemeris for the X-ray eclipses from a 10 year series of \chandra ~and \xmm ~observations, it was found that the published radial velocity data of \cite{SF} are offset by 1/4 of an orbit from the geometric alignment implied by the eclipses. The phase shift is such that the periodic doppler shift of the He [$\lambda$4686] line does not pass though zero at mid-eclipse. Instead, it is the maximum blueshift that occurs at mid-eclipse which is expected to coincide with inferior conjunction of the WR primary.  

Variations in He II line-width are evident from Table~2 of  \cite{SF}, which suggests the $FWHM$ is largest when the RV is closest to zero, and smallest at maximum redshift. In \cite{vanKerkwijk1996} the line profile model consists of a weak broad contribution from the hot mostly ionized stellar wind, plus a narrow velocity-modulated line from the cool gas in the shadow which alternately passes in front of and behind the star, attaining its minimum width (and greatest strength) at maximum red-shift. $v_{wind}$ in the line-forming region is substantially lower than the terminal wind velocity.  
\cite{Vilhu2009} used Chandra HETG spectra to map the ionization and velocity fields of the wind in Cyg X-3, largely confirming the picture developed by \cite{vanKerkwijk1996} and \cite{Hanson2000} and showing that the wind is highly ionized out to many times the binary separation. The X-ray FeXXVI line RV curve appears to trace the compact object's motion, with the required phase relationship to the IR emission and absorption features, resulting in a mass function consistent with that obtained by \cite{Hanson2000}.

The appeal of the \cite{vanKerkwijk1996} model is that it isolates a single phenomenon (photoionization) that seems likely to dominate interaction in IC 10 X-1;  it is not unique in producing phase-shifts and newer models for Cyg X-3 successfully reproduce many elements of {\it its} more complex dynamics. The phenomena encompassed by those models probably apply to a lesser extent in IC10 X-1 since the orbital separation in Cyg X-3 is smaller, resulting in the compact object being engulfed by the stellar envelope. The flow of material in that system is influenced by both ionization and tidal interactions, leading to density structures such as plumes, wake-trails, and shocks; while IC10 X-1 has so far been observed in a state analogous to the quiescent state of Cyg X-3.

\subsection{Spectral Analysis of IC10 X-1}
IC10 is distant and so grating spectra have not yet been obtained. \cite{Wang2005} fit the \xmm~EPIC (pulse height) spectrum with a multicolor disk blackbody model appropriate to a BH in the low hard state. Finding an inner accretion disk temperature of $T_{in}$=1.11$_{-0.05}^{+0.06}$ keV at a radius of $R_{in}$=25$_{-6}^{+4}$ km, disk inclination angle $>$ 57$^{\circ}$, subsolar metal abundance and $N_H \sim$10$^{22}$\nh. The point source is blended with an extended diffuse structure of effective radius of $50_{-7}^{+23}$ km, which appears to be an ionized X-ray nebula surrounding the compact object. Unfortunately, these early observations did not accumulate enough counts during eclipse to fully isolate the spectral components. By looking at the variation in hardness ratio (HR) during eclipse egress in the \chandra~data, \cite{Laycock2015} found evidence of energy dependence and noted that the hardness ratio increases {\it before} the flux begins its climb at egress.  Energy dependence is not expected in the eclipse profile of an X-ray point source passing behind a fully ionized absorber. The modulation would be dominated by free electron scattering which is energy independent over this wavelength range. At the more moderate ionization fraction expected deep in the core of the wind (and especially in the shadow of the WR star), single-electron ions of e.g. O, C, Si, and Fe will be present (as seen by \cite{Vilhu2009} in Cyg X-3) and contribute significant soft X-ray absorption. 

A much longer \xmm~EPIC observation analyzed by \cite{Barnard2014} reveals evidence for an accretion-disk corona. Their Fig.~2 shows a clear HR modulation during eclipse. The interpretation is that different parts of the emission region (disk and corona) become covered by the absorber as the eclipse progresses. They avoid the word eclipse in favour of ``dip" because the passage of the compact object through (and behind) the wind gives rise to varying amounts of scattering and absorption at all phases. Thus, what we call the eclipse is referred to by them as the ``main dip."  The inner-edge temperature for their preferred accretion-disk spectral fit (Model V) is $T_{in}$=1.68 keV, and this component disappears during eclipse, while about 10\% of the spatially-extended, comptonized corona component remains visible (``partial covering"). We note that Model V is not the best-fit model of \cite{Barnard2014}. Model III with $T_{in}$=0.2 keV and a compact corona was a better fit, but was rejected (in part) because it does not conform to the assumption of a (massive) stellar mass BH. The very large number of parameters required in these models results from the lack of information about the spectrum above 10 keV, and the inability of current datasets to resolve the soft X-ray lines expected from hydrogen-like heavy ions in the absorbing core of the wind. 

In what follows, we adopt the model shown in Figure~\ref{fig:map} for optical line emission during X-ray eclipses and in \S~2 we make the case for a low-mass compact object in IC10 X-1, possibly a NS although a low-mass BH cannot be ruled out. In \S~3, we discuss the implications for other HMXBs and ULXs, and in \S~4 we summarize our conclusions. 

\section[]{A Low-Mass Compact Object}
If we in fact do not know the mass function of IC10 X-1, then the only hard evidence on which to base a model for the system comes from the orbital period and the stellar parameters of the WR primary. The X-ray luminosity and eclipse duration are informative but they are subject to many more parameters (wind density, ionization, accretion efficiency, magnetic field), and consequently they can be used to support competing non-unique solutions. Here we analyze the plausibility of a much lower mass BH or a NS as the compact object. 

The mass of the WR star is reasonably well constrained by the optical classification of \cite{Clark2004} with a most probable value of $M_{WR}$=35\msun ~and a range of 17-35 \msun. For a neutron star companion, a commonly accepted range for the mass is $M_{NS} = 1.26 - 2.5$ \msun, where the lower bound is the Chandrasekhar limit for a white dwarf minus the binding energy lost in collapse to a NS, and the upper bound is the Tolman-Oppenheimer-Volkoff (TOV) limit for a NS. The highest reliable measurement is 2.0 \msun ~and current EOS allow 0.1-2.9 \msun. For BHs there are no physical bounds to the mass and any dynamically confirmed object significantly in excess of 2.5 \msun ~is regarded as a probable BH.

Applying Kepler's law with NS masses yields an orbital separation of $(a_1 + a_2) =  14$\rsun ~for the lowest mass combination (1.4 + 17\msun), and 19\rsun ~for the highest mass combination  (2.5 + 35\msun). Naturally, the NS orbit shrinks now to 14-19\rsun ~relative to the 19-23\rsun ~range used in prior investigations that assumed a 23-32 \msun ~BH (see Table~\ref{tab:calculations}). At these closer separations, the stellar wind density encountered by the compact object increases and the primary approaches filling its Roche lobe. The size of the Roche lobe of the primary for all of these combinations of WR star and NS masses is $r_{1}\approx 0.6(a_1 + a_2) = 8$-11\rsun.

\begin{table}
\caption{Orbital Parameters for a NS or a BH in IC10 X-1}
  \label{tab:calculations}
  \begin{center}
    \leavevmode
\begin{tabular}{llllllll} 
\hline
$M_{WR}$ & $M_X$         & $a$     &  $v_{WR}$   & $v_X$            & R(s)        & $R_{WR}$   & $T_{WR}$  \\
\msun    &   \msun       &    \rsun            &   km/s      & km/s                     &   \rsun     & \rsun           &      hr                 \\
\hline
 35        & 1.4                       & 18.99             &  23.96          & 599.10                    & 8.06        & 2              & 1.24   \\
 35        & 2.5                       & 19.18             &  41.95          & 587.32                    & 8.14        & 2              & 1.23     \\
 17        & 1.4                       & 15.13             & 37.76           & 458.57                     & 6.42        & 1.5           & 1.17 \\
 17        & 2.5                       & 15.42             &  64.88          & 441.15                     & 6.55        & 1.5           & 1.15  \\
\hline
35         & 32                       & 23.27             & 364.7          & 398.9                           & 9.88        & 2               &    1.01 \\ 
17         & 23                         & 19.59             & 369.7          & 273.3                         & 8.32        & 1.5            &   0.90   \\ 

\hline
\end{tabular}
\end{center}
Radius of shadowing body R(s) determined from observed eclipse duration $T_{min}$=5 hr. Inferred duration for eclipse by the WR star is $T_{WR}$. Table assumes inclination $i=90^{\circ}$
\end{table}

Analyses of the \chandra ~and \xmm ~eclipse profiles by \cite{Laycock2015} and \cite{Barnard2014}, respectively, measured slightly different durations for minimum flux (5 hr vs. 5.2 hr), and ingress/egress phases ($\sim$1 hr vs. $\sim$2 hr). This appears to be due to the broader and more gradual eclipse seen by \xmm. Differences in the method used to estimate the diameter of the WR star also led to small variations in the estimates of the size of the eclipsing body (4 vs. 5 for $R_{WR}$) . \cite{Barnard2014} estimated the effective stellar radius $R_{WR}$ seen by the X-rays to be the blackbody radius implied by the color temperature reported by \cite{Clark2004}; while we used the tabulated WR mass-radius relation of \cite{Langer1989} which produces slightly smaller results.

\begin{figure}
  \begin{center}
    \leavevmode
      \includegraphics[trim=0 1cm 0 1cm, clip, angle=0,width=9 cm]{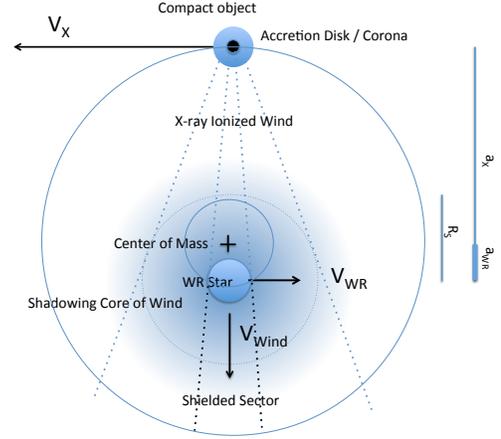}
       \caption{Schematic diagram of the IC10 X-1 system for $M_{WR}$ = 35 \rsun ~and $M_X$ = 1.4-2.5 \msun ~with all components drawn approximately to scale (See Table~1). X-ray eclipses occur as the WR star's dense wind core region passes through the line of sight to the compact object. At mid-eclipse, a ``shielded sector" of the wind that lies in the shadow of the WR star will be pointing directly toward the observer. The vector $v_{wind}$ represents the velocity of the partially ionized material that gives rise to the optical He II emission line. Enhanced absorption and scattering by this cooler material contributes to the depth and shape of the eclipse minimum. The cross represents the center of mass of the binary. The large mass ratio causes the compact object to move in a wide orbit at high speed while the WR star moves only slightly by comparison. For small $M_X$, the center of mass lies inside the WR envelope.}
     \label{fig:map}
  \end{center}
  \vspace{-0.5cm}
\end{figure}

In the case of a NS secondary, the orbital separation does not shrink enough to cast doubt on the results of \cite{Barnard2014} and \cite{Laycock2015} in regards to the general size of the shadowing absorber and the extended corona. The binary separation and the orbital velocities change together, so that the sizes inferred from eclipse timing remain relatively unchanged; Table~\ref{tab:calculations} shows a contraction of about 20\%. The eclipse is still several times larger than can be explained by an eclipse of the WR star.  Hence, the X-ray flux modulation is still required to be dominated by absorption and scattering in the dense core of the stellar wind, perhaps in the cooler shadowed material (Figure~\ref{fig:map}. 

The parameter that changes most dramatically is the orbital velocity of the WR star $v_{WR}$, which decreases by a factor of $\sim$10-15 for $M_{X} = 1.4$ \msun ~(Table~\ref{tab:calculations}). An important consequence is that now $v_{wind} >> v_{WR}$ which is a requirement for the \cite{vanKerkwijk1996} model to be valid.  Then the observed RV curve is expected to reflect only a small advance in phase due to the orbital and rotational velocities of the WR star that are both very small compared to $v_{wind}$. The complete picture is more complicated: at increasing distances $r$ from the WR star, the shielded sector contains material that was ejected earlier from the point of view of an observer corotating in the shadow (Fig~\ref{fig:map}). So, as the wind expands outward, its tangential velocity (fixed at the stellar surface value $v_{rot}$) is rapidly outpaced by the motion of the shadow whose velocity goes as $r\omega$. Thus the projected velocity of the emission line will be influenced by the phase advance provided by the stellar surface velocity, and the trailing which becomes larger at higher altitudes. The result will be a line profile that depends on the viewing angle and hence on orbital phase. A constant centre of mass or $\gamma$ velocity is expected in the RV curve, this is not evident in \cite{SF} who measured their line centroids relative to nebular lines, which probably co-move with the binary. In Cyg X-3, $\gamma$ has not been measured in optical/IR spectra despite that system having a large X-ray derived value, a situation \cite{Hanson2000} attribute to turbulence.   

Model calculations for IC 10 X-1 by \cite{Tutukov2013} found that for accepted values of the stellar wind parameters in WR stars, the X-ray source should have been more luminous by at least one order of magnitude. A similar conclusion was reached by \cite{Barnard2014}, that a mere 2\% of the Bondi accretion rate is sufficient to power the observed X-rays. Both of these calculations assumed large BH masses; however, an adopted lower mass (permitted by our reinterpretation of the dynamical data) eliminates the discrepancy. Specifically, for typical NS parameters ($M_X=1.4$\msun~ and $R_X=10$~km)
and for the measured dynamical parameters ($M_{WR}=35$\msun, $P_{orbit}=1.45$~d, and $v_{wind}=370$~km~s$^{-1}$), we find from eq.~(4) of \cite{Stella1986} 
for direct wind accretion that $L_X = 1.5\times 10^{38}$~\lx~
when we also adopt the value of $\dot{M}=2.6\times 10^{-7}$~\msun yr$^{-1}$ proposed by \cite{Tutukov2013} for the rate of the mass captured by the accretor.

\section{Discussion}
As a putative 24 \msun ~BH with a 35 \msun ~companion located in a low metallicity ($Z$=0.3 Z$_{\sun}$) galaxy, IC 10 X-1 has contributed to the picture of metallicity as determining maximum stellar BH mass. Dynamically confirmed BHs are few in number, so it is striking that the most massive, including ULX BHs, and intermediate-mass (IMBH) candidates (e.g. HLX-1 in ESO 243-49: \citealt{Farrell2014}) have been found in galaxies with extremely low $Z$. Detailed stellar evolution calculations \citep{Belc2010} show that metallicity determines the mass-loss rate in the progenitor star's wind and hence the remnant mass. The feasibility of IMBHs growing in cluster or galaxy cores from stellar BH `seeds'  via dynamical interactions is critically sensitive to the initial stellar BH mass. The above is no longer supported by IC10 X-1 if the compact object has a low mass.
 
The debate about the nature of the compact objects in ULXs (e.g. \citealt{Gilfanov2004}, \citealt{Luangtip2014}) 
continues to this day as new observations arise, and the balance has tipped recently in favour of stellar-mass BHs and NSs
(M101 ULX-1, \citealt{Liu2013}; NGC7793 P13, \citealt{Motch2014}; M-82 X-2, \citealt{Bachetti2014}) rather than IMBHs.
It is actually interesting that the observations do not favour the ``simpler'' resolution of this issue:
the main difficulty with ULXs is that they should not radiate above the Eddington limit 
($L_{\rm Edd}=1.3\times 10^{38}(M_X/M_\odot)$~erg~s$^{-1}$), and IMBHs
with masses $M_X$$\sim$10$^{2-4}$\msun ~would remain comfortably below this limit. 


A preliminary attempt to find pulsating ULXs (other than M82 X-2) in {\it XMM-Newton} archival data \citep{Doroshenko2014} found none, but more sensitive surveys will follow. Our attention has turned to IC10 X-1 as a NS candidate because of
its modest (for a ULX) luminosity ($L_X\leq 1.2\times 10^{38}~{\rm erg~s}^{-1}$) 
and its phase-shifted RV curve \citep{Laycock2015}. In such a case,
the Corbet diagram \citep{Corbet} predicts the NS pulsations to lie in the \pspin $<$1~s range for the observed \porb$=$1.45~d. Notwithstanding the absence of any other known WR+NS binary, we suppose that such a system will lie between the SG and RLOF systems.
 
The lack of pulsations could plausibly be due to limitations of the current datasets. Time resolution in typical imaging mode for \chandra ~ACIS is 3.4~s, for \xmm ~EPIC-MOS 2.6~s and EPIC-PN 73~ms. The recent discovery of a 1.37~s pulsar in M82 X-2 by \cite{Bachetti2014} with {\it NuSTAR} shows that such objects can hide in plain sight if their spin periods lie below the timing threshold.  At very short orbital periods (such as \porb = 1.45~d for IC10 X-1 and \porb = 2.5~d for M82 X-2), pulsations suffer doppler blurring because of the orbital motion of the NS during an observation. 
The time required for the path length variation to accumulate a light-crossing time of \pspin/2 and hence to go out of phase is $\delta$t = c \pspin / (2$v_{orbit})$, where $c$ is the speed of light. This is as little as 250~s for the maximum projected velocity in Table~\ref{tab:calculations} of $v_{orbit}=$600 km/s (Doppler shift $z$=0.004). At the average uneclipsed \xmm ~PN count-rate of 0.5 c/s, it will be challenging to detect such pulsations.

Stellar winds can wash out high-frequency variability when photons reaching the observer scatter off electrons lying at a wide range of distances from the NS \citep{Kylafis1987}. When the light-travel time across the scattering region is comparable to the pulse period, the modulation is lost. 
Assuming the scattering core of the WR wind to have a radius of $\leq$8 \rsun (Table~\ref{tab:calculations}), pulse smearing would be significant for \pspin$<$18~s. 

A NS with a magnetic field in the usual range for young objects ($10^{12}$~G) should be a pulsar at IC 10 X-1's accretion rate. The existence of a slow 7 mHz QPO \citep{Pasham2013} could then be attributed to the interaction of the magnetic field with the inner edge of the accretion disk. The type of QPO seen in IC10 X-1 is not BH-specific and occurs in many NS systems \citep{Bozzo2009}. It has long been known that high B-field NSs can switch off their pulsations yet continue to emit X-rays. The best known example is A0538-66 which has been detected on numerous occasions, but only once have pulsations been seen \citep{Skinner1982}. Magnetospheric accretion that is gated by rotational inhibition may also be able to create unpulsed emission. This mechanism has previously been invoked to explain the persistent low luminosity of some NS+Be systems in quiescence and as a way to turn off accretion in SFXTs where high mass-transfer rates are continuously present.   


\section{Conclusions}
Based on a finding that the RV curve of the WR X-ray binary IC10 X-1 is not phase-aligned with the X-ray eclipses, we have presented the case that the compact object in this system, rather than 
being a 20-30\msun ~BH,  is a NS or a lower-mass BH. If the RV curve is a projection of the WR's wind velocity, then there is no longer any {\it dynamical evidence} for a massive BH in IC10 X-1. 
In this letter we have focussed on application of the \cite{vanKerkwijk1996} model for the interaction between the radiation field and the stellar wind in HMXBs, and we are keenly aware that the true picture is more complex. The wind velocity profile and spatial ionization structure remain unconstrained by observation, requiring higher resolution optical/IR spectra as input for modeling. The shape of the X-ray modulation is likely related to the scattering and absorption properties of the gas in the shielded sector which crosses the line of sight at mid-eclipse. Hard X-ray observations could probe the effects of scattering on the eclipse profile, while motivation for phase-resolved X-ray line spectra is established by \cite{Vilhu2009}. 
The observed X-ray luminosity of X-1 (\citealt{Wang2005}) lies within the range exhibited by both NSs and stellar-mass BHs in HMXBs although it is substantially higher than most persistent NS+SG systems, and closer to the Be systems at the peaks of their giant outbursts. Super-Eddington luminosity has been seen in A0538-66 and recently M82 X-2, so the possibility exists for a NS in IC 10 X-1.  
The system is similar in many regards to other supposed BH-IMXBs which now should be examined closely with regard to the possibility that this phenomenon is more widespread. 

This work was supported in part by NASA grant NNX14-AF77G.

\label{lastpage}

\end{document}